%% file: acl_latex.tex
\pdfoutput=1

\documentclass[11pt]{article}

\usepackage[final]{acl}

\usepackage{times}
\usepackage{latexsym}
\usepackage{amsfonts}
\usepackage{amsmath}
\usepackage{multirow}
\usepackage{colortbl}
\usepackage{tabularx}

\usepackage[T1]{fontenc}

\usepackage[utf8]{inputenc}

\usepackage{microtype}

\usepackage{inconsolata}

\usepackage{graphicx}
\usepackage{xcolor}

\definecolor{Gray}{gray}{0.9}

\title{Improving Retrieval in Sponsored Search by Leveraging Query Context Signals}

\author{Akash Kumar Mohankumar$^{\dagger}$\hspace{0.2cm} Gururaj K$^{\dagger}$ \\\ \bf{Gagan Madan}$^{\dagger}$ \hspace{0.2cm} \bf{Amit Singh}$^{\dagger}$  \\
  $^\dagger$Microsoft, India  \\
  {\tt \{makashkumar,gururajk,gaganmadan,siamit\}@microsoft.com} 
}

\begin{document}
\maketitle

\input{latex/abstract}
\input{latex/introduction}
\input{latex/proposed_work}
\input{latex/results}
\input{latex/conclusion}

\bibliography{custom}

\appendix
\input{latex/appendix}

\end{document}

%% file: latex/abstract.tex
\begin{abstract}
Accurately retrieving relevant bid keywords for user queries is critical in Sponsored Search but remains challenging, particularly for short, ambiguous queries. Existing dense and generative retrieval models often fail to capture nuanced user intent in these cases. To address this, we propose an approach to enhance query understanding by augmenting queries with rich contextual signals derived from web search results and large language models, stored in an online cache. Specifically, we use web search titles and snippets to ground queries in real-world information and utilize GPT-4 to generate query rewrites and explanations that clarify user intent. These signals are efficiently integrated through a Fusion-in-Decoder based Unity architecture, enabling both dense and generative retrieval with serving costs on par with traditional context-free models. To address scenarios where context is unavailable in the cache, we introduce \textit{context glancing}, a curriculum learning strategy that improves model robustness and performance even without contextual signals during inference. Extensive offline experiments demonstrate that our context-aware approach substantially outperforms context-free models. Furthermore, online A/B testing on a prominent search engine across 160+ countries shows significant improvements in user engagement and revenue. 
\end{abstract}

%% file: latex/introduction.tex
\section{Introduction}
\input{latex/tables/main_example}
Sponsored search is a primary revenue model for many search engines, where advertisements are displayed alongside organic search results. In this model, advertisers bid on specific keywords to target user intents relevant to their business objectives. They can select various match types to determine how closely their keywords align with user queries. For instance, the exact match type restricts keyword matching to queries that precisely share the same intended meaning. In contrast, more flexible match types, such as phrase and smart match, enable advertisers to target a broader range of search intents with their keywords. Retrieving relevant bid keywords for a user query is a critical task in sponsored search, directly influencing both the revenue generated and the quality of ads served to users. \\

\textbf{Prior Work and Challenges:} Traditionally, keyword retrieval has been approached as a standard information retrieval task. Some methods utilize dual encoders to map queries and keywords into a shared semantic space, with optional use of a one-vs-all classifier to rerank the shortlisted keywords \cite{decaf, SiameseXML, ngame}. Another line of research considers this problem as a constrained Natural Language Generation (NLG) task, where language models transform queries into relevant keywords \cite{cloverv1, baidu_generative_trie, ProphetNetAds, pixar}. A recent approach, Unity \cite{unity}, integrates Dense Retrieval (DR) and NLG methodologies into a unified model, harnessing the strengths of both while requiring only a single model. Despite these advancements, existing methods struggle with short and ambiguous queries. As shown in Table \ref{tab:main_example}, the query \textit{"ad623armz reel"} - which actually refers to an integrated circuit by Analog Devices - is incorrectly associated with keywords related to fishing reels, leading to the retrieval of irrelevant ads. This issue largely stems from the use of shallow transformer models, which are necessary to meet strict online latency requirements but have limited capacity to encode complex world knowledge. Consequently, understanding specialized terms like \textit{"ad623armz"} becomes challenging without any additional information. \\

\textbf{Our Contribution:} To address these limitations, we introduce {Augmented Unity}, a framework for context-aware retrieval in sponsored search. Our approach utilizes a large dynamic cache to enrich queries with contextual signals. This cache includes organic search results, such as titles and snippets from the top-k web documents for each query. Additionally, we create a \textit{Query Profile} for each query, containing multiple rewrites and a description of the user's potential intent, generated using GPT-4, and store them in the cache. If a query is absent from the cache, an offline pipeline is triggered to generate this context, ensuring its availability for future instances of the same query within a short timeframe. Our Augmented Unity model employs a Fusion-in-Decoder architecture, which enables  efficient processing of diverse contexts. We train this model using a curriculum learning strategy termed \textit{context glancing}, which progressively introduces more challenging scenarios with varying levels of context availability. Our evaluations show that {Augmented Unity} significantly outperforms the context-free Unity model by 19.9\% in exact match Precision at 100, while maintaining a comparable online GPU serving cost (within 7-9\%). Moreover, {Augmented Unity}, trained with context glancing, demonstrates robust performance even when context signals are absent, matching the performance of the Unity model. Through extensive online A/B testing, we show that {Augmented Unity} achieves a 1\% and 1.4\% increase in ad revenue for English and non-English queries, respectively, without any statistical change in ad defects.

%% file: latex/tables/main_example.tex
\begin{table}[]
\centering
\def\arraystretch{1.0}%
\resizebox{1\linewidth}{!}{
\begin{tabular}{l}
\hline
\textbf{Query}: ad623armz reel                                                                                                                                                                                                                                                                                                                                                                                                                                            \\ \hline
\begin{tabular}[c]{@{}l@{}}\textbf{Web Results}:     \\ \textbf{Title 1}: AD623ARMZ-\textcolor[rgb]{0.133, 0.549, 0.133}{REEL7} \textcolor[rgb]{0.133, 0.549, 0.133}{Analog Devices Inc. | Integrated} \\ \textcolor[rgb]{0.133, 0.549, 0.133}{Circuits (ICs)} |   DigiKey \vspace{0.1cm} \\\textbf{Snippet 1}: AD623ARMZ-\textcolor[rgb]{0.133, 0.549, 0.133}{REEL7} – Instrumentation Amplifier\\ 1 Circuit Rail-to-Rail 8-MSOP from \textcolor[rgb]{0.133, 0.549, 0.133}{Analog Devices Inc} …\end{tabular}                                                                                                                                                                                            \\ \hline
\begin{tabular}[c]{@{}l@{}}\textbf{Query Profile}:\\ \textbf{Rewrites}:  ad623armz analog amplifier reel, \textcolor[rgb]{0.133, 0.549, 0.133}{analog devices} reel \\of ad623 armz, ad623armz tape and reel, ... \vspace{0.1cm} \\ \textbf{Intent}: The user is looking for a specific \textcolor[rgb]{0.133, 0.549, 0.133}{integrated circuit (IC)}\\ \textcolor[rgb]{0.133, 0.549, 0.133}{chip}, the AD623ARMZ, that is sold in a reel package. A reel\\ package is a type of bulk packaging that contains  many IC chips ...\end{tabular} \\ \hline
\begin{tabular}[c]{@{}l@{}}\textbf{Retrieved Keywords}: \vspace{0.15cm}\\   \textbf{Unity (Context-free)}:\\ 1. \textcolor{red}{fishing rods} reels and gear\\ 2. \textcolor{red}{fishing} reels with \textcolor{red}{rod}\\ 3. \textcolor{red}{fishing rods fishing} reels \vspace{0.15cm}  \\  \textbf{Augmented Unity (with Context)}: \\ 1. ad623armz reel \textcolor[rgb]{0.133, 0.549, 0.133}{analog devices}\\ 2. ad623armz \textcolor[rgb]{0.133, 0.549, 0.133}{reel7 analog devices}\\3. ad623armz \textcolor[rgb]{0.133, 0.549, 0.133}{microchip}\end{tabular}                                                                  \\ \hline
\end{tabular}}
\caption{Table illustrates how incorporating web results and LLM-generated Query Profile enables our proposed Augmented Unity model to retrieve relevant keywords.}
\label{tab:main_example}
\end{table}

%% file: latex/proposed_work.tex
\section{Proposed Method: Augmented Unity}
\begin{figure*}
    \centering
\centering
\includegraphics[width=0.99\textwidth]{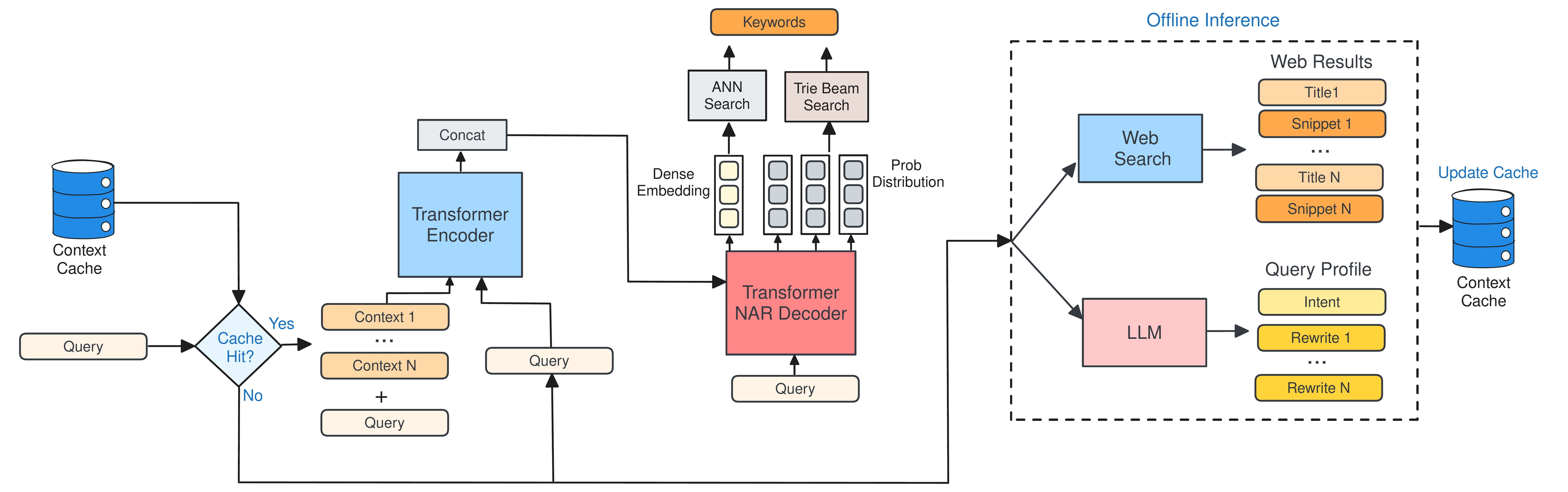}
    \caption{Overview of the Augmented Unity architecture for context-aware retrieval. The system leverages a context cache to store and retrieve pre-computed query context signals, employing an offline pipeline for cache misses. The Augmented Unity model, utilizing a Fusion-in-Decoder approach, effectively combines query representations}
    \label{fig:main_diagram}
\end{figure*}

Figure \ref{fig:main_diagram} provides an overview of our Augmented Unity workflow. We use a context cache to store and retrieve query context signals. For incoming queries, we first check the cache. In case of a cache miss, an asynchronous offline pipeline is triggered to generate the context signals. These signals are then used to update the cache, ensuring their availability for future occurrences of the same query. This section is structured as follows: Section \ref{subsec:quey_context_signals} details the various context signals employed. Section \ref{subsec:model_architecture_training} outlines our efficient model architecture for integrating these signals. Section \ref{subsec:context_glancing} introduces \textit{context glancing}, our curriculum learning strategy designed to enhance model performance and robustness in scenarios with missing context signals. 


\subsection{Query Context Signals}
\label{subsec:quey_context_signals}
We utilize two sources of query-level signals to provide richer context and disambiguate user intent: 

{\flushleft \bf Web Search:} Organic search results, often highly relevant to user intent, provide valuable contextual information. We utilize the title and snippet of each web result, offering a concise summary of the webpage content. For example, as demonstrated in Table \ref{tab:main_example}, web titles and snippets help identify \textit{"ad623armz"} as an electronic IC chip from Analog Devices Inc., leading to the retrieval of keywords with spans such as \textit{"analog devices"} and \textit{"microchip"}. We mine web results from the logs of a prominent search engine. To account for location-based variations in search results, we utilize the country where a query is most frequently searched. In cases of multiple occurrences, the most recent result is prioritized to ensure up-to-date information. We cache the top 10 web results per query and periodically refresh them to incorporate new queries and update existing ones.

{\flushleft \bf LLM-generated Query Profile:} We leverage the reasoning abilities and extensive "world knowledge" of large language models like GPT-4 to generate Query Profile. These profiles comprise of query rephrases and explanations of potential user intents, aiding in disambiguation. Table \ref{tab:main_example} illustrates this where Query Profile includes clarifying rewrites like \textit{"ad623armz analog amplifier reel"} and \textit{"analog devices reel of ad623 armz"}. Further, it includes a concise explanation of the possible user intent, offering crucial background information for query understanding. 

We utilize various query normalization techniques, including spell correction, to map minor query variations to a canonical form for cache storage and lookups. With these normalizations, our cache hit rate is approximately 70\% of all user search requests.

\subsection{Model Architecture \& Training}
\label{subsec:model_architecture_training}
We now discuss our proposed architecture that effectively combines the aforementioned context signals with the original query. Inspired by the Unity framework \cite{unity}, we use a shared model to perform both dense and non-autoregressive (NAR) generative retrieval, leveraging the complementary benefits of the two approaches at the cost of one. Our model consists of a encoder $\mathcal{E}$ with $L_e$ transformer encoder layers and a NAR decoder $\mathcal{D}$ with $L_d$ transformer decoder layers. We first encode the query $Q = \{{w}^0_1, \dots, {w}^0_{l^0}\}$ of length $l^0$ and each context signal ${C}^{i} = \{{w}^i_1, \dots, {w}^i_{l^i}\}$ of length $l^i$, independently using the encoder, obtaining hidden states $\mathbf{H} = \{\mathbf{H}^{i}\}_{i=0}^{n}$, where $\mathbf{H}^{i} = \{\mathbf{h}^i_1, \dots, \mathbf{h}^i_{l^i}\} \in \mathbb{R}^{l^i \times d}$,  $\mathbf{H}^{0}$ corresponds to the query's hidden states, $\{\mathbf{H}^{i}\}_{i=1}^{n}$ represents the hidden states of the $n$ context signals, and $d$ represents the hidden size. Following the Fusion-in-Decoder (FiD) approach \cite{fid}, we concatenate these hidden states into $\mathbf{\tilde{H}} = [\mathbf{H}^0, \dots, \mathbf{H}^n] \in \mathbb{R}^{l \times d}$, where $l = \sum_{i=0}^{n} l_i$, and leverage them within our NAR decoder $\mathcal{D}$. This decoder utilizes bidirectional attention without a causal mask and receives the original query $Q$ as input, as opposed to right-shifted target tokens in autoregressive models. After processing through the $L_d$ decoder layers, we obtain final hidden states $\mathbf{G} = \{\mathbf{g}_1, \dots, \mathbf{g}_{l^0}\} \in \mathbb{R}^{l^0 \times d}$.  These are used to compute the dense retrieval embedding $\mathbf{e}(Q)$ and the NAR token probabilities $P(k_t | Q)$:
\begin{align}
    \nonumber \mathbf{e}(Q) &= \mbox{Attention}(\Tilde{\mathbf{g}}, \mathbf{G}\mathbf{W}^{K}, \mathbf{G}\mathbf{W}^{V}) \\
    \nonumber P(k_t | Q) &= \mbox{Softmax}(\mathbf{W}^{O}\mathbf{g}_t)
\end{align}
where $\mathbf{W}^{K}, \mathbf{W}^{V} \in \mathbb{R}^{d \times d'}$ are attention key and value matrices, $\Tilde{\mathbf{g}} \in \mathbb{R}^{d'}$ is a learnable query vector, $\mathbf{W}^{O} \in \mathbb{R}^{V \times d}$ represents the language modeling head's weight matrix, and $d'$ and $V$ correspond to the dense embedding and vocabulary sizes, respectively. We train the model using a combination of the contrastive loss with in-batch negatives for DR and the negative log-likelihood loss for NLG:

\begin{align}
    \nonumber \mathcal{L}(\theta, \mathcal{B}) &= \frac{-1}{|\mathcal{B}|}(\mathcal{L}^{N}(\theta, \mathcal{B}) + \lambda \mathcal{L}^{D}(\theta, \mathcal{B})) \\
    \nonumber \nonumber \mathcal{L}^{N}(\theta, \mathcal{B}) &= \sum_{Q,K \in \mathcal{B}} \sum_{k_t \in K} \log P (k_t | Q) \\
    \nonumber \mathcal{L}^{D}(\theta, \mathcal{B}) &= \sum_{Q,K \in \mathcal{B}} \log \frac{\exp(\mbox{Sim}(Q, K))}{\sum_{K' \in \mathcal{B}} \exp(\mbox{Sim}(Q, K'))}
\end{align}

where $\mathcal{B}$ represents a training batch of query-keyword pairs, $\theta$ represents the model's learnable parameters, $\lambda$ is a hyperparamter for weighting the two losses, and $\mbox{Sim}(Q, K)$ is the cosine similarity between their dense embeddings: $\frac{\mathbf{e}(Q)^T\mathbf{e}(K)}{||\mathbf{e}(Q)|| \cdot ||\mathbf{e}(K)||}$. 


\subsection{Context Glancing}
\label{subsec:context_glancing}
A key challenge in our proposed workflow is ensuring the retrieval model functions effectively both with and without available context. To address this, we introduce \textit{context glancing}, a curriculum learning based strategy that gradually accustoms the model to scenarios where context might be absent. Initially, the model is trained for several epochs with context provided for all training examples. Subsequently, we progressively drop context from a subset of training examples, increasing the drop rate throughout the training process. We employ a combination of random and structured context dropping methods. For $d_{rand}$ fraction of queries, we randomly drop $k$ contexts, with $k$ sampled uniformly from 1 to $n$. For $d_{web}$ and $d_{qp}$ fractions of queries, we drop all web-based and query profile-based contexts, respectively. Finally, for $d_{all}$ fraction of queries, we remove all context signals. All drop rate parameters ($d_{rand}$, $d_{web}$, $d_{qp}$, and $d_{all}$) are gradually increased during training based on a linear schedule following an initial warm-up phase with no context dropping. Our approach follows the principles of curriculum learning by gradually introducing increasingly difficult scenarios. The model initially learns in a context-rich environment, becoming progressively accustomed to handling cases with partial or even complete absence of context signals. Notably, the Augmented Unity architecture adapts to varying numbers of input contexts. 

%% file: latex/results.tex
\section{Results \& Discussion}
In this section, we begin by outlining the offline experimental setup, including the datasets used and evaluation metrics in Section \ref{subsec:experimental_setup}. Sections \ref{subsec:offline_results} and \ref{subsec:gpt4_evaluation} discuss the core offline results, showcasing the performance gains achieved by Augmented Unity. We further dissect the impact of different components within our approach through detailed ablation studies in Section \ref{subsec:ablation_study}. Finally, in Section \ref{subsec:online_results}, we discuss our online experiments and results.  

\input{latex/tables/main_results}
\subsection{Experimental Setup}
\label{subsec:experimental_setup}
{\flushleft \bf Dataset:} We construct a dataset of high-quality query-keyword pairs extracted from the search logs of a prominent search engine, encompassing 40 languages globally. The training set consists of approximately 60M unique queries, 240M unique keywords, and 900M query-keyword pairs. The keywords are chosen to be either exact, phrase, or smart match variants of the query. Our test set consists of 1M queries sampled across all languages. Retrieval is performed against a corpus of 1B keywords sampled from the full bid keyword corpus.

{\flushleft \bf Evaluation Metrics:} Evaluating the quality of retrieved keywords for a given query often necessitates nuanced understanding beyond simple n-gram matching \cite{unity}. While recent studies \cite{AnnoLLM} demonstrate the superior accuracy of LLMs like GPT-4 compared to crowd-sourced human annotators for this task, using such models for large-scale evaluation over billions of query-keyword pairs remains computationally expensive. To address this, we first curate a large-scale dataset annotated with query-keyword match type quality (exact, phrase, smart) using GPT-4. We then use this dataset to train a smaller, computationally efficient student model capable of accurately predicting match type quality. We utilize this student model's predictions to compute Precision@K, separately for each match type. 

{\flushleft \bf Baselines:} We compare Augmented Unity against several competitive context-free baselines: CLOVERv2 \cite{unity}, PIXAR \cite{pixar}, SimCSE \cite{simcse}, NGAME \cite{ngame}, and Unity \cite{unity}. Due to space constraints, we provide further details on the baselines and the implementation details in Appendix 
 \ref{sec:appendix_experimental_setup}. 

\subsection{Offline Results}
\label{subsec:offline_results}
Table \ref{tab:main_results} presents a comparative analysis of Augmented Unity's retrieval performance against prominent context-free NLG and DR methods. Our results demonstrate that Augmented Unity consistently outperforms the best-performing context-free baselines across all match types. Specifically, the NLG component of Augmented Unity surpasses Unity by 12-20\% and the state-of-the-art PIXAR method by 5-13\% in P@100. Note that our approach of leveraging query context signals is complementary to the idea of scaling up the vocabulary in PIXAR. Similarly, the DR component of Augmented Unity showcases substantial gains, with a 7-20\% improvement in P@100 compared to the baseline Unity DR model. These findings underscore the effectiveness of our approach in leveraging additional context to improve query understanding. Further, despite processing 13x more tokens due to the inclusion of query context, the GPU serving cost of Augmented Unity remains comparable to Unity, within 7-9\%. This efficiency stems from our use of the Fusion-in-Decoder architecture, where inference complexity scales as  $O(NL_{max}^2)$, in contrast to $O(N^2L_{max}^2)$  when concatenating and encoding all $N$ contexts together ($L_{max}$  is the max sequence length of the contexts).

\subsection{Evaluation with GPT-4 as Judge}
\label{subsec:gpt4_evaluation}
\input{latex/tables/gpt4_labelled_results}
We also conducted additional evaluations with GPT-4 as the evaluation model. We randomly sampled 1000 queries each from English, French, and German from our test set, and retrieved the top 50 keywords from Unity and Augmented Unity. These query-keyword pairs were then evaluated by GPT-4, which provided binary judgments for exact, phrase, and smart match quality. Table \ref{tab:gpt4_as_judge} presents the results, demonstrating that Augmented Unity consistently outperforms Unity across all three languages and match types. We observe an average relative improvement of 12.6\% for NLG and 17.7\% for DR, further validating the effectiveness of our proposed approach in improving keywords retrieval. 

\subsection{Ablation Studies}
\label{subsec:ablation_study}
\input{latex/tables/ablation_context}
Augmented Unity incorporates three key components: (i) leveraging various query contexts from both web search results and LLM-generated Query Profiles, (ii) utilizing multiple instances of each context type, (iii) using context glancing to enhance model robustness to scenarios with missing context signals. To understand the contribution of each component, we conducted a series of ablation studies, which are detailed below:

{\flushleft \bf Context Type:} Augmented Unity leverages four distinct types of query context: web titles, web snippets, Query Profile rewrites, and Query Profile intent. Table \ref{tab:ablation_context} shows the impact of using these different context types on the performance of Augmented Unity DR, as measured by P@100. The results reveal several key insights: (1) Irrespective of the type, incorporating any context leads to a substantial performance gain over the context-free scenario. This highlights the inherent value of each signal in enhancing query understanding. (2) Among the four types, Query Profile intent yields the largest performance gains. This could be because the intent derived from GPT-4 often provides a concise and accurate explanation of the user's underlying intent, directly aiding in disambiguation. (3) Utilizing both web results and Query Profile context outperforms using either source alone. This indicates that these sources provide complementary information, highlighting the importance of leveraging them jointly.

{\flushleft \bf Number of Context Instances:} While utilizing various context types proves beneficial, determining the optimal number of contexts to use is crucial. To investigate this, we varied the number of web titles, web snippets, and query profile rewrites. Table \ref{tab:ablation_num_context} displays the P@100 scores for different number of contexts used per type. As evident from the results, increasing the number of contexts per type from 1 to 4 consistently enhances retrieval performance. However, further increasing the number of contexts to 10 leads to a performance decline. This suggests that while incorporating multiple contexts per type can be advantageous up to a certain point, including an excessive number of potentially noisy contexts can negatively impact retrieval accuracy.

\input{latex/tables/ablation_num_context}
\input{latex/tables/ablation_context_glancing}
{\flushleft \bf Context Glancing:} Table \ref{tab:ablation_context_glancing} demonstrates the effectiveness of our context glancing strategy in enhancing model robustness. Without context glancing, the model exhibits a significant performance drop of 27.5\% (in terms of phrase match P@100) when context signals are unavailable during inference, highlighting an over-reliance on availability of context. However, incorporating context glancing consistently improves performance across all scenarios, regardless of context availability.  For instance, in the complete absence of context, context glancing leads to a 20.8\% improvement in phrase match P@100.  Remarkably, even when full context is provided, context glancing still yields a 5.3\% performance gain. This suggests that our approach of gradually exposing the model to increasingly challenging scenarios not only enhances robustness to missing context but also leads to a more generalizable model with improved overall performance. 

\subsection{Online A/B Testing}
\label{subsec:online_results}
\input{latex/tables/online_ab_results}
\input{latex/tables/decile_wise_results}
To validate the effectiveness of Augmented Unity in a real-world setting, we conducted extensive online A/B testing for 30 days on live traffic of a prominent commercial search engine, spanning over 160 countries. We deployed both the NLG and DR components of Augmented Unity and compared their performance against an ensemble of state-of-the-art retrieval techniques, including proprietary DR and NLG models, Unity, large language models, extreme classification, and graph-based methods. We measured the overall revenue, ad clicks, Quick Back Rate, and ad defect. Quick Back Rate (QBR) denotes the percentage of ad clicks with users quickly returning to the search results page. Ad defect, measured by offline relevance models, denotes the percentage of irrelevant ads shown to users. Table \ref{tab:online_results} summarizes the online A/B testing results, segmented by English and non-English queries. Augmented Unity improved user engagement, yielding statistically significant increases in overall clicks -- a 0.72\% lift for non-English queries and a 0.33\% lift for English queries. Critically, these gains were not accompanied by any statistically significant degradation in QBR or ad defect. This suggests that Augmented Unity was able to retrieve keywords that aligned with the user intent.

We also analyzed our online A/B experiment results by grouping queries into frequency-based deciles. Decile 1 contains highly frequent queries, while decile 10 consists of a large number of rare queries. Table \ref{tab:decile_wise_stats} shows the query coverage (the fraction of queries for which any sponsored content was shown) and ad impressions (the total number of ads displayed). We observe an increase in both query coverage and ad impressions across all deciles, with particularly strong gains on tail queries, which are often longer and more ambiguous.  As a result of the improved user engagement, we observed substantial revenue gains – 1.43\% for non-English queries and 1.02\% for English queries – underscoring the tangible business impact of our approach within a real-world sponsored search ecosystem.

%% file: latex/tables/main_results.tex
\begin{table*}[]
\centering
\def\arraystretch{1.1}%
\resizebox{0.8\textwidth}{!}{
\begin{tabular}{lccccccc}
\hline
\multicolumn{1}{c|}{\multirow{2}{*}{\textbf{Model}}} & \multicolumn{2}{c|}{{Exact Match}}                                         & \multicolumn{2}{c|}{{Phrase Match}}                                          & \multicolumn{2}{c|}{{Smart Match}}                                          & \multirow{2}{*}{\textbf{\begin{tabular}[c]{@{}c@{}}GPU\\ Cost\end{tabular}}} \\ \cline{2-7}
\multicolumn{1}{c|}{}                                & \multicolumn{1}{c}{{P@100}} & \multicolumn{1}{c|}{{P@200}} & \multicolumn{1}{c}{{P@100}} & \multicolumn{1}{c|}{{P@200}} & \multicolumn{1}{c}{{P@100}} & \multicolumn{1}{c|}{{P@200}} &                                                                              \\ \hline
\multicolumn{8}{c}{{NLG}}                                                                                                                                                                                                                                                                                                                                       \\ \hline
\multicolumn{1}{l|}{CLOVERv2}                        & \multicolumn{1}{c}{6.78}           & \multicolumn{1}{c|}{4.97}           & \multicolumn{1}{c}{16.58}          & \multicolumn{1}{c|}{14.08}          & \multicolumn{1}{c}{31.83}          & \multicolumn{1}{c|}{28.04}          & 1.17x                                                                        \\ 
\multicolumn{1}{l|}{Unity NLG}                       & \multicolumn{1}{c}{7.00}           & \multicolumn{1}{c|}{5.14}           & \multicolumn{1}{c}{16.93}          & \multicolumn{1}{c|}{14.36}          & \multicolumn{1}{c}{32.69}          & \multicolumn{1}{c|}{28.81}          & 1.17x                                                                        \\ 
\multicolumn{1}{l|}{PIXAR}                           & \multicolumn{1}{c}{7.50}           & \multicolumn{1}{c|}{5.48}           & \multicolumn{1}{c}{18.12}          & \multicolumn{1}{c|}{15.30}          & \multicolumn{1}{c}{34.87}          & \multicolumn{1}{c|}{30.54}          & 1.41x                                                                        \\ 
\rowcolor{Gray} \multicolumn{1}{l|}{Aug Unity NLG (Ours)}      & \multicolumn{1}{c}{\textbf{8.43}}  & \multicolumn{1}{c|}{\textbf{6.21}}  & \multicolumn{1}{c}{\textbf{19.06}} & \multicolumn{1}{c|}{\textbf{16.26}} & \multicolumn{1}{c}{\textbf{36.88}} & \multicolumn{1}{c|}{\textbf{32.35}} & 1.26x                                                                        \\ \hline
\multicolumn{8}{c}{{DR}}                                                                                                                                                                                                                                                                                                                                        \\ \hline
\multicolumn{1}{l|}{NGAME}                           & \multicolumn{1}{c}{9.66}           & \multicolumn{1}{c|}{6.77}           & \multicolumn{1}{c}{16.21}          & \multicolumn{1}{c|}{12.48}          & \multicolumn{1}{c}{42.86}          & \multicolumn{1}{c|}{39.47}          & 1.00x                                                                        \\ 
\multicolumn{1}{l|}{SimCSE}                          & \multicolumn{1}{c}{9.70}           & \multicolumn{1}{c|}{6.82}           & \multicolumn{1}{c}{16.50}          & \multicolumn{1}{c|}{12.78}          & \multicolumn{1}{c}{43.88}          & \multicolumn{1}{c|}{40.26}          & 1.00x                                                                        \\ 
\multicolumn{1}{l|}{Unity DR}                        & \multicolumn{1}{c}{10.58}          & \multicolumn{1}{c|}{7.52}           & \multicolumn{1}{c}{18.34}          & \multicolumn{1}{c|}{14.20}          & \multicolumn{1}{c}{48.29}          & \multicolumn{1}{c|}{44.30}          & 1.00x                                                                        \\ 
\rowcolor{Gray} \multicolumn{1}{l|}{Aug Unity DR (Ours)}       & \multicolumn{1}{c}{\textbf{12.74}} & \multicolumn{1}{c|}{\textbf{8.87}}  & \multicolumn{1}{c}{\textbf{20.86}} & \multicolumn{1}{c|}{\textbf{15.82}} & \multicolumn{1}{c}{\textbf{52.05}} & \multicolumn{1}{c|}{\textbf{45.66}} & 1.09x                                                                        \\ \hline
\end{tabular}}
\caption{\textbf{Performance and Efficiency Comparison of Augmented Unity with Context-Free Methods}. Precision (P) at 100 and 200 for different match types are reported, along with relative GPU serving cost compared to NGAME}
\label{tab:main_results}
\end{table*}

%% file: latex/tables/gpt4_labelled_results.tex
\begin{table}[]
\centering
\def\arraystretch{1.1}%
\resizebox{1\linewidth}{!}{
\begin{tabular}{lcccccc}
\hline
\multicolumn{1}{c|}{\multirow{2}{*}{Lang}} & \multicolumn{2}{c|}{EM}                                    & \multicolumn{2}{c|}{PM}                                     & \multicolumn{2}{c}{SM}                \\ \cline{2-7} 
\multicolumn{1}{c|}{}                      & \multicolumn{1}{c}{Unity} & \multicolumn{1}{c|}{Aug} & \multicolumn{1}{c}{Unity} & \multicolumn{1}{c|}{Aug} & \multicolumn{1}{c}{Unity} & Aug \\ \hline
\multicolumn{7}{c}{NLG}                                                                                                                                                                                        \\ \hline
\multicolumn{1}{l|}{English}               & \multicolumn{1}{c}{12.05}  & \multicolumn{1}{c|}{14.29}      & \multicolumn{1}{c}{24.77} & \multicolumn{1}{c|}{26.74}     & \multicolumn{1}{c}{44.28} & 48.40     \\ 
\multicolumn{1}{l|}{French}                & \multicolumn{1}{c}{9.91}  & \multicolumn{1}{c|}{11.90}      & \multicolumn{1}{c}{25.27} & \multicolumn{1}{c|}{27.43}     & \multicolumn{1}{c}{47.33} & 51.95     \\ 
\multicolumn{1}{l|}{German}                & \multicolumn{1}{c}{9.90}  & \multicolumn{1}{c|}{11.93}      & \multicolumn{1}{c}{24.38} & \multicolumn{1}{c|}{26.43}     & \multicolumn{1}{c}{41.39} & 45.54     \\ \hline
\multicolumn{7}{c}{DR}                                                                                                                                                                                        \\ \hline
\multicolumn{1}{l|}{English}               & \multicolumn{1}{c}{19.09} & \multicolumn{1}{c|}{23.21}     & \multicolumn{1}{c}{28.44} & \multicolumn{1}{c|}{32.99}     & \multicolumn{1}{c}{58.15} & 66.23     \\ 
\multicolumn{1}{l|}{French}                & \multicolumn{1}{c}{17.55} & \multicolumn{1}{c|}{21.75}     & \multicolumn{1}{c}{29.28} & \multicolumn{1}{c|}{34.69}     & \multicolumn{1}{c}{61.86} & 69.65     \\ 
\multicolumn{1}{l|}{German}                & \multicolumn{1}{c}{17.34} & \multicolumn{1}{c|}{21.07}     & \multicolumn{1}{c}{29.18} & \multicolumn{1}{c|}{34.36}     & \multicolumn{1}{c}{58.31} & 66.32     \\ \hline
\end{tabular}}
\caption{\textbf{GPT-4 as Judge}: Precision@50 for Unity and Augmented Unity with GPT-4 as the evaluation model}
\label{tab:gpt4_as_judge}
\end{table}

%% file: latex/tables/ablation_context.tex
\begin{table}[]
\centering
\def\arraystretch{1.1}%
\resizebox{0.99\linewidth}{!}{
\begin{tabular}{l|c|c|c}
\hline
\textbf{Context} (Num)           & {EM} & {PM} & {SM} \\ \hline
None (0)                       & 10.45        & 17.35       & 45.70       \\ \hline
Web Title (4)                 & 11.61        & 19.13       & 49.58       \\ 
Web Snippet (4)               & 11.16        & 18.45       & 48.07       \\ 
QProfile Rewrites (4)          & 11.60        & 19.32       & 49.47       \\ 
QProfile Intent (1)           & 12.07        & 20.06       & 49.86       \\ \hline
Web Title (4) + Snippet (4)        & 11.63        & 19.18       & 49.70       \\ 
QProfile Rewrites (4) + Intent (1) & 12.17        & 20.15       & 49.96       \\ \hline
All (13)                       & \textbf{12.74}        & \textbf{20.86}       & \textbf{52.05}       \\ \hline
\end{tabular}}
\caption{Precision@100 for Augmented Unity DR with different types of query contexts used during inference}
\label{tab:ablation_context}
\end{table}

%% file: latex/tables/ablation_num_context.tex
\begin{table}[]
\centering
\def\arraystretch{1.0}%
\resizebox{0.99\linewidth}{!}{
\begin{tabular}{c|c|c|c|c}
\hline
\textbf{\# Context} & \textbf{Context}                                                                      & {EM} & {PM} & {SM} \\ \hline
4                   & \begin{tabular}[c]{@{}c@{}}1 Title + 1 Snippet + \\ 1 Rewrite + Intent\end{tabular}   & 12.03        & 19.89       & 49.20       \\ \hline
7                   & \begin{tabular}[c]{@{}c@{}}2 Title + 2 Snippet +\\ 2 Rewrite + Intent\end{tabular}    & 12.32        & 20.08       & 50.04       \\ \hline
13                  & \begin{tabular}[c]{@{}c@{}}4 Title + 4 Snippet +\\ 4 Rewrite + Intent\end{tabular}    & \textbf{12.74}        & \textbf{20.86}       & \textbf{52.05}       \\ \hline
31                  & \begin{tabular}[c]{@{}c@{}}10 Title + 10 Snippet +\\ 10 Rewrite + Intent\end{tabular} & 12.55        & 20.39       & 50.57       \\ \hline
\end{tabular}}
\caption{Precision@100 for Augmented Unity DR with different numbers of context instances used}
\label{tab:ablation_num_context}
\end{table}

%% file: latex/tables/ablation_context_glancing.tex
\begin{table}[]
\centering
\def\arraystretch{1.1}%
\resizebox{0.95\linewidth}{!}{
\begin{tabular}{l|c|c|c}
\hline
\multicolumn{1}{c|}{\textbf{Context}} & \textbf{w/o GG} & \textbf{w CG} & \textbf{$\Delta$} \\ \hline
None                                   & 14.36           & 17.35         & 20.8\%         \\ \hline
Web Title                              & 17.04           & 19.13         & 12.3\%         \\ 
Web Snippet                            & 16.17           & 18.45         & 14.1\%         \\ 
QProfile Rewrites                      & 17.06           & 19.32         & 13.3\%         \\ 
Qprofile Intent                        & 19.30           & 20.15         & 4.4\%          \\ \hline
Web Title + Snippet                    & 17.14           & 19.18         & 11.9\%         \\ 
Qprofile Rewrites + Intent             & 19.30           & 20.15         & 4.4\%          \\ \hline
All                                    & 19.80           & 20.86         & 5.3\%          \\ \hline
\end{tabular}}
\caption{Phrase Match P@100 for Augmented Unity DR trained with and without Context Glancing (CG)}
\label{tab:ablation_context_glancing}
\end{table}

%% file: latex/tables/online_ab_results.tex
\begin{table}[]
    \centering
    \begingroup
    \def\arraystretch{1}
    \resizebox{0.95\linewidth}{!}{
    \begin{tabular}{l|c|c|c|c}
    \hline
    {Language} & {$\Delta$ Revenue} & {$\Delta$ Clicks} & {$\Delta$ QBR} & {$\Delta$ Defect} \\ \hline
    English                   & \textbf{{1.00\%}}           & \textbf{{0.33\%}}            & \textcolor{gray}{0.01\%}  & \textcolor{gray}{0.02\%}      \\ \hline
    Non-english               & \textbf{{1.44\%}}           & \textbf{{0.72\%}}          & \textcolor{gray}{0.18\%} & \textcolor{gray}{0.19\%}      \\ \hline
\end{tabular}}
\caption{Online A/B results on a commercial search engine. Gray color indicates p-value > 0.01}
\label{tab:online_results}
\endgroup
\end{table}
    

%% file: latex/tables/decile_wise_results.tex
\begin{table}[]
\centering
\def\arraystretch{1.1}%
\resizebox{1\linewidth}{!}{
\begin{tabular}{lcccc}
\hline
\multicolumn{1}{c|}{\multirow{2}{*}{Decile}} & \multicolumn{2}{c|}{$\Delta$ Query Coverage}                                    & \multicolumn{2}{c}{$\Delta$ Ad Impressions} \\ \cline{2-5} 
\multicolumn{1}{c|}{}                      & \multicolumn{1}{c}{English} & \multicolumn{1}{c|}{Non-English} & \multicolumn{1}{c}{English} & Non-English  \\ \hline
\multicolumn{1}{l|}{1}               & \multicolumn{1}{c}{0.66\%}  & \multicolumn{1}{c|}{1.49\%}      & 0.84\% & 1.71\%     \\ 
\multicolumn{1}{l|}{2}               & \multicolumn{1}{c}{0.09\%}  & \multicolumn{1}{c|}{0.70\%}      & 0.01\% & 0.35\%     \\ 
\multicolumn{1}{l|}{3}               & \multicolumn{1}{c}{1.16\%}  & \multicolumn{1}{c|}{0.48\%}      & 0.37\% & 0.86\%     \\ 
\multicolumn{1}{l|}{4}               & \multicolumn{1}{c}{0.49\%}  & \multicolumn{1}{c|}{0.51\%}      & 0.26\% & 0.48\%     \\ 
\multicolumn{1}{l|}{5}               & \multicolumn{1}{c}{0.41\%}  & \multicolumn{1}{c|}{0.51\%}      & 0.59\% & 0.61\%     \\ 
\multicolumn{1}{l|}{6}               & \multicolumn{1}{c}{0.35\%}  & \multicolumn{1}{c|}{0.56\%}      & 0.55\% & 0.70\%     \\ 
\multicolumn{1}{l|}{7}               & \multicolumn{1}{c}{0.61\%}  & \multicolumn{1}{c|}{0.71\%}      & 0.66\% & 0.54\%     \\ 
\multicolumn{1}{l|}{8}               & \multicolumn{1}{c}{0.76\%}  & \multicolumn{1}{c|}{0.69\%}      & 0.98\% & 0.74\%     \\ 
\multicolumn{1}{l|}{9}               & \multicolumn{1}{c}{0.51\%}  & \multicolumn{1}{c|}{1.12\%}      & 1.47\% & 1.09\%     \\ 
\multicolumn{1}{l|}{10}               & \multicolumn{1}{c}{1.07\%}  & \multicolumn{1}{c|}{0.76\%}      & 1.66\% & 0.76\%     \\ 
\hline
\end{tabular}}
\caption{Percentage Change in query coverage and Ad impression for different deciles in online A/B tests on sponsored search}
\label{tab:decile_wise_stats}
\end{table}

%% file: latex/conclusion.tex
\section{Conclusion}
In this paper, we introduced Augmented Unity, a novel approach that leverages rich query context to enhance sponsored search retrieval. By integrating web search results and GPT-4 generated Query Profiles, Augmented Unity effectively disambiguates user intent and retrieves more relevant bid keywords. Furthermore, our proposed context glancing strategy ensures robust performance even when contextual information is unavailable. Through extensive offline experiments and rigorous online A/B testing on a commercial search engine, we  showed substantial improvements in key metrics such as retrieval accuracy, user engagement (ad clicks), and ad revenue. These findings underscore the significant potential of incorporating contextual information for achieving more effective and efficient sponsored search retrieval.

%% file: latex/appendix.tex
\newpage
\section{More Details on Experimental Setup}
\label{sec:appendix_experimental_setup}
This section provides further details about the experimental setup, including baseline descriptions and implementation details.

\subsection{Baselines}
We compare Augmented Unity against strong baselines in both dense retrieval (DR) and natural language generation (NLG) for keyword retrieval:

{\flushleft \bf Dense Retrieval:} We utilize the same baselines as reported in \cite{unity}: NGAME \cite{ngame} and SimCSE \cite{simcse}. Both NGAME and SimCSE employ a siamese dual encoder architecture to represent queries and keywords in a dense vector space. However, they differ in their approaches to curating negatives and their training objectives. SimCSE uses a contrastive InfoNCE-style loss with in-batch random negatives, while NGAME adopts a triplet margin loss. Moreover, NGAME uses a clustering-based strategy to curate batches, ensuring that the batches themselves contain hard negatives. In addition to these methods, we also compare against the DR component of the Unity model, which utilizes a dual encoder architecture similar to NGAME. 

{\flushleft \bf NLG:} We compare our model against CLOVERv2, the NLG component of Unity \cite{unity}, and PIXAR \cite{pixar} as NLG baselines. All these models are non-autoregressive, predicting the keyword token distribution independently and in parallel. Autoregressive models, though potentially effective, are impractical for online deployment due to significantly higher latency and inference costs \cite{unity}. CLOVERv2 utilizes an encoder-based architecture with a language modeling head to map the hidden states to the vocabulary space, thereby obtaining the final token distributions. It then performs a constrained beam search using a Trie to produce the predicted keywords. Unity leverages a similar approach while sharing its model with the DR component. PIXAR scales up the target vocabulary in non-autoregressive models to include phrases.

\subsection{Implementation Details}
{\flushleft \bf Model details:} For Augmented Unity, We employ a 4-layer encoder and 4-layer decoder architecture ($L_e=4, L_d=4$) with a hidden size ($d$) of 512, utilizing the multilingual 250k vocabulary of XLM-R \cite{xlmr}. The model is trained from scratch. For NGAME, SimCSE, Unity, and PIXAR, we utilize the XLM-R base encoder model initialized with pretrained weights. The dense retrieval embedding size ($d'$) is set to 128.

{\flushleft \bf Training details:} All models are trained with a learning rate of $5\times10^{-5}$, 1000 warmup steps, and an effective batch size of 16384 for 10 epochs. When using context glancing, we train for first 3 epoch without any context dropping and then linearly increase each dropping rate parameter ($d_{rand}$, $d_{web}$, $d_{qp}$, and $d_{all}$) to a maximum value of 10\%. We utilize the Adam optimizer with a linear decay learning rate scheduler. The dense retrieval loss weight hyperparameter ($\lambda$) is set to 1. We train all our models on 16x AMD Mi200 GPUs with DeepSpeed stage 1. 

\section{Multilingual Analysis}
\label{sec:appendix_multilingual_analysis}
\input{latex/tables/language_analysis}
Table \ref{tab:result_language_wise} presents a language-wise breakdown of phrase match P@100 for Augmented Unity DR and Unity DR.  Augmented Unity demonstrates consistent improvements across all languages, including high-resource languages like English, German, and French. Notably, we observe a substantial 18.5\% relative gain in phrase match P@100 for English. The gains are even more pronounced for low-resource languages like Maltese and Icelandic, highlighting the benefits of incorporating additional context, particularly when training data is limited. These results suggest that incorporating additional context is particularly beneficial for languages with limited training data. We also note that the Query Profile intent is generated in English regardless of the query language and hence provides valuable cross-lingual context, aiding in improved query understanding for tail languages. 

\section{Error Analysis}
\label{sec:error_analysis}
\input{latex/tables/error_analysis}

While Augmented Unity leads to significantly better model performance overall, there are a few cases where adding context can lead to errors. Table \ref{tab:error_analysis} shows a few selected failure modes for the model. 

\section{Alternatives to GPT-4 for Query Profile}
\label{subsec:slm_results}
\input{latex/tables/finetune_slm}
Running GPT-4 for a large number of queries to build a sufficiently large Query Profile cache can get fairly expensive. In order to mitigate this, we experiment with a finetuning based approach with Mistral 7B \cite{Mistral7B} with LoRA \cite{hu2021lora} for generating QProfile Rewrites. We evaluate the generated rewrites on three metrics precision, novelty and diversity. Precision is calculated as described in \ref{subsec:gpt4_evaluation}, while novelty and diversity are calculated by GPT-4, where we get a score between 1-5 for novelty for each of rewrites and a score between 1-5 for diversity for all 10 rewrites together. Table \ref{tab:finetuned_mistral} shows a comparison of results between rewrites generated by GPT-4 and Mistral-7b. Our results show that fine-tuned Mistral-7B achieves comparable performance to GPT-4 in terms of query rewrite precision, at the cost of novelty and diversity, while significantly reducing inference costs. Improving novelty and diversity without losing out on precision could be a fruitful direction for a future work.

%% file: latex/tables/language_analysis.tex
\begin{table}[]
\centering
\def\arraystretch{1.0}%
\resizebox{0.7\linewidth}{!}{
\begin{tabular}{l|c|c|c}
\hline
Language   & Unity & Aug   & $\Delta$   \\ \hline
English    & 22.69 & 26.88 & 18.5\% \\ 
German    & 23.00 & 27.88 & 21.2\% \\ 
French    & 22.83 & 27.45 & 20.2\% \\ \hline
Maltese    & 5.86  & 8.44  & 44.1\% \\ 
Icelandic  & 7.62  & 10.81 & 41.8\% \\ 
Lithuanian & 9.55  & 13.53 & 41.8\% \\ 
Latvian    & 10.38 & 14.46 & 39.3\% \\ 
Albanian   & 7.95  & 10.93 & 37.5\% \\ 
Slovenian  & 11.58 & 15.63 & 35.0\% \\  \hline
\end{tabular}}
\caption{Phrase Match P@100 for Augmented Unity and Unity DR across different languages}
\label{tab:result_language_wise}
\end{table}

%% file: latex/tables/error_analysis.tex
\begin{table*}[]
\centering
{
\begin{flushleft}
Example: \textbf{Noise from Web Snippets}
\end{flushleft}
\begin{tabular}{p{0.2\linewidth} | p{0.7\linewidth}}
\hline
Query & \texttt{compensation injury claim} \\ \hline
Web Snippet & \texttt{Report an occupational injury to the Compensation  Fund Report injuries sustained by your employees  in the course of their work to the Compensation Fund within seven days of the occurrence. The fund  covers permanent, casual workers, trainees and apprentices who are injured in the course of their work and lose income or are impaired as a result} \\ \hline
\multicolumn{2}{c}{{Predictions}} \\ \hline
Augmented Unity & \texttt{what to claim on income tax} \\ \hline
Unity & \texttt{compensation for injury} \\ \hline
\end{tabular}
\begin{tabular}{p{0.9\linewidth}}
\multicolumn{1}{c}{{Analysis}} \\ \hline
The model incorrectly picked up the token “income” from the context, leading to an irrelevant keyword related to income tax. \\ \hline
\end{tabular}
\linebreak
\linebreak
\begin{flushleft}
Example: \textbf{Location Bias}
\end{flushleft}
\begin{tabular}{p{0.2\linewidth} | p{0.7\linewidth}}
\hline
Query & \texttt{cruises singles} \\ \hline
Web Title & \texttt{Singles Cruise Deals | Marella Cruises | TUI.co.uk} \\ \hline
\multicolumn{2}{c}{{Predictions}} \\ \hline
Augmented Unity & \texttt{cruises holidays uk} \\ \hline
Unity & \texttt{cruises for singles only} \\ \hline
\end{tabular}
\begin{tabular}{p{0.9\linewidth}}
\multicolumn{1}{c}{{Analysis}} \\ \hline
The cached web result was location-specific (UK), leading to a geographically biased keyword retrieval. \\ \hline
\end{tabular}
}
\caption{\textbf{Some qualitative examples of failure modes for Augmented Unity compared to Unity.}}
\label{tab:error_analysis}
\end{table*}

%% file: latex/tables/finetune_slm.tex
\begin{table*}[]
\centering
\def\arraystretch{1}%
\resizebox{0.65\linewidth}{!}{
\begin{tabular}{c|c|c|c}
\hline
Model & Precision@10 & \begin{tabular}[c]{@{}c@{}}Avg Novelty\\ (1-5)\end{tabular} & \begin{tabular}[c]{@{}c@{}}Avg Diversity\\ (1-5)\end{tabular} \\ \hline
GPT-4 & 6.9 & 2.19 & 2.53 \\ \hline
Finetuned Mistral-7b & 7.62 & 2.05 & 2.36 \\ \hline
\end{tabular}
}
\caption{Performance of Finetuned Mistral-7b vs GPT-4 for QProfile Rewrites}
\label{tab:finetuned_mistral}
\end{table*}